\documentclass[prd,aps,onecolumn,12pt]{revtex4}

\usepackage{amsmath}

\begin{document}
\title{Measuring $D^0-\overline D^0$ mixing in $D^0(\overline D^0)\to f_0(980) K^{*}$ and more}
\author{ Wei Wang$^{a}$ and Cai-Dian L\"u$^{a,b}$ }
\affiliation{
 $^a$ Institute of High Energy Physics, Chinese Academy
of Sciences, Beijing 100049, P.R. China\\
 $^b$  Theoretical Physics Center for Science Facilities,  Chinese Academy
of Sciences, Beijing 100049, P.R. China}

\begin{abstract}
We investigate the $D^0-\overline D^0$ mixing through the doubly
Cabibbo suppressed (DCS) channel $D^0\to f_0(980)K^{*0}$ and its
charge conjugate channel, in which the $K^{*0}$ meson is
reconstructed in both $K^+\pi^-$ and $K_S\pi^0$ final state.
Although the decay $D^0\to f_0(980)K^{*}$ has a small branching
ratio, the final state mesons are relatively easy to identify. The
$f_0(980)$ meson can be replaced by  $\pi^+\pi^-$ in which
$\pi^+\pi^-$ form an $S$-wave state, or a longitudinally polarized
vector meson $\rho^0,\omega$. All mixing parameters, including the
mass difference and decay width difference, can be extracted by
studying the time-dependent decay width of these channels. We  show
that the method is valid in all regions for mixing parameters and it
does not depend on the strong phase difference.

\end{abstract}

\maketitle
%

The neutral $D^0$ meson mixes with its CP-conjugate $\overline D^0$
through  box diagrams in the standard model (SM). In the box
diagrams, the $b$ quark coupling is small and the $s$ and $d$ quark
have small masses. Thus the mixing of $D^0-\overline D^0$ in the SM
is very small, measurements on the mixing parameters serve as a
probe to detect new physics   scenarios. In the presence of
mixtures, the interference in  transition amplitudes of $D$ and
$\overline D^0$ is possible. Experimentalists can study the
time-dependent decay widths to extract the mixing parameters.
Recently, the BaBar, Belle and CDF collaborations have reported
their measurements on $D^0-\bar D^0$
mixing~\cite{Aubert:2007wf,Staric:2007dt,Abe:2007rd,:2007uc,:2007en}.
Meanwhile, a number of theoretical methods are proposed to measure
the mixing
parameters~\cite{Gronau:2001nr,Rosner:2003yk,Asner:2005wf,Grossman:2006jg,Li:2006wx,Cheng:2007uj,Xing:2007sd,Sinha:2007ck}.

In this note, we propose a new method to extract the mixing
parameters in $D^0-\overline D^0$ mixing: using the  doubly Cabibbo
suppressed (DCS) channel $D^0\to M K^{*0}$ ($M$ denotes a $f_0(980)$
meson, a non-resonant $S$-wave $\pi^+\pi^-$ state, or a
longitudinally polarized $\rho^0$ or $\omega$ meson) and its
CP-conjugate channel. The decay $D^0\to MK^{*0}$ can proceed in term
of the DCS amplitude or the amplitude from the mixing followed by a
Cabibbo allowed decay $\overline D^0\to M K^{*0}$. The two
amplitudes interfere with each other. If the $K^{*0}(\bar K^{*0})$
meson is reconstructed in both $K^+\pi^-(K^-\pi^+)$ and $K_S\pi^0$
final state, the mixing parameters can be determined through the
time-dependent studies on the decay widths.



At present, there is no direct experimental measurement on $D^0\to
f_0(980) \bar K^{*0}$. The branching ratio for $D^0\to f_0(980)K_S$
is given by~\cite{Yao:2006px}:
\begin{eqnarray}
 {\cal BR}(D^0\to f_0(980) K_S)=(1.36^{+0.30}_{-0.22})\times
 10^{-3},\label{eq:experimentalf0KS}
\end{eqnarray}
where $f_0(980)$ is identified in the $\pi^+\pi^-$ final state. This
value can be used to estimate the decay rate of  $D^0\to f_0(980)
\bar K^{*0}$, together with the following results~\cite{Yao:2006px}:
\begin{eqnarray}
 {\cal BR}(D^0\to \pi^0 K_S)=(1.14\pm0.12)\%,\nonumber\\
 {\cal BR}(D^0\to \pi^0\bar K^{*0}_{K^-\pi^+})=(1.91\pm0.24)\%,\nonumber\\
 {\cal BR}(D^0\to \pi^0\bar K^{*0}_{K_S\pi^0 })=(6.3^{+1.8}_{-1.5})\times
 10^{-3},\label{eq:experimentalpi0K0}
\end{eqnarray}
where the subscript denotes the daughter mesons to reconstruct the
$\bar K^{*0}$ meson. Under the factorization assumption, the decay
amplitudes are proportional to $D^0\to \pi^0(f_0)$ form factor times
the $K_0(K^{*0})$ decay constants. From
Eqs.~\eqref{eq:experimentalf0KS} and \eqref{eq:experimentalpi0K0},
we expect a somewhat larger branching fraction for the $D^0\to
f_0(980)\bar K^{*0}$ decay channel than ${\cal BR}(D^0\to
f_0(980)K_S)$. Since the sum of masses of $f_0(980)$ and $K^{*0}$ is
close to the mass of $D^0$ meson, this decay channel will be
suppressed by the phase space. To obtain some information on the
phase space suppression, we can compare the two decays $D^0\to\phi
K_S$ and $D^0\to\phi \bar K^{*0}$~\cite{Yao:2006px}:
\begin{eqnarray}
 {\cal BR}(D^0\to \phi_{K^+K^-}  K_S)=(2.10\pm0.16)\times
 10^{-3},\nonumber\\
 {\cal BR}(D^0\to \phi_{ K^+K^-}  \bar K^{*0}_{K^-\pi^+})=(1.01\pm0.20)\times
 10^{-4}.\label{eq:experimentalphi0K0}
\end{eqnarray}
The branching ratio of $D^0\to\phi \bar K^{*0}$ is about $5\%$ of
that of $D^0\to\phi K_S$. Assume the same suppression for $D^0\to
f_0(980)\bar K^{*0}$, we expect this decay has a branching ratio of
${\cal O}(10^{-5})$ which can be studied by the future experiments.
In our method, one can also replace the $f_0(980)$ by the
non-resonant $\pi^+\pi^-$ final state where $\pi^+\pi^-$ is
restricted to be an $S$-wave state. The three-body (non-resonant)
decay $\bar D^0\to \pi^+\pi^-\bar K^{*0}$ also possesses a sizable
branching fraction~\cite{Yao:2006px}:
\begin{eqnarray}
 {\cal BR}( D^0\to \pi^+\pi^- \bar K^{*0}_{K^-\pi^+})=(9.7\pm2.1)\times
 10^{-3}.
\end{eqnarray}
In the following, we will take $D\to f_0(980)K^*$ as an
example and use $f_0$ to denote $f_0(980)$ for convenience.

For $D^0\to f_0 K^{*0}(\bar K^{*0})$, the transition in quark level
is either $c\to s\overline d u$ or $c\to d\overline s u$. The former
transition is proportional to $V_{cs}V_{ud}^*\sim 1$ and the latter
is suppressed by the CKM matrix elements: $V_{cd}V_{us}^*\sim 0.04$.
Since there is only one amplitude for each decay in the SM, the
direct CP asymmetries are 0 and the amplitudes satisfy the relation:
$A(D^0\to f_0 K^{*0})=A(\overline D^0\to f_0\overline K^{*0})$ and
$A(D^0\to f_0\overline K^{*0})=A(\overline D^0\to f_0 K^{*0})\equiv
A_{f_0K^*}$. Neglecting direct CP asymmetries, one often defines the
two parameters $r_{f_0K^*}$ and $\delta_{f_0K^*}$ by:
\begin{eqnarray}
 -r_{f_0K^*} e^{-i\delta _{f_0K^*}} \equiv \frac{A(D^0\to f_0 K^{*0})}{A(D^0\to f_0\overline K^{*0})}
 = \frac{A(\overline D^0\to f_0\overline K^{*0})}{A(\overline D^0\to
 f_0K^{*0})}.\label{eq:ratio}
\end{eqnarray}

Under the assumption of CPT invariance,   mass eigenstates of
neutral $D$ meson system are given by:
\begin{eqnarray}
 |D_1\rangle= p|D^0\rangle + q |\overline D^0\rangle,\;\;\;\; |D_2\rangle=p|D^0\rangle - q |\overline
 D^0\rangle,
\end{eqnarray}
where $p$ and $q$ satisfy the normalization condition
$|p|^2+|q|^2=1$.  $\phi$ is defined as the phase  of $q/p$: $\phi=
{\rm arg}(q/p)$. In the SM, the phase $\phi$ is very small and thus
the large value for this parameter will definitely imply the
presence of new physics. $D_1$ and $D_2$ have different masses and
different decay widths. The differences are defined as: $\Delta
M=M_1-M_2$ and $\Delta \Gamma=\Gamma_1-\Gamma_2$. For convenience,
experimentalists also use the two parameters $ x\equiv \frac{\Delta
M}{\Gamma}$, $y\equiv \frac{\Delta \Gamma}{2\Gamma},$ where $\Gamma$
is the averaged decay width of the $D$-mesons.

In the limit of $x\ll 1$, $y\ll1$ and $\Gamma t\ll1$, the
time-dependent decay amplitudes squared of $D^0\to f$ in which there
is a purely $D^0$ state at $t=0$ is given by:
\begin{eqnarray}
 |A(D^0(t)\to f)|^2 &=& e^{-\Gamma t} \left[ X_f +Y_f\Gamma t+ Z_f (\Gamma
 t)^2+...\right],\label{eq:time-dependentDtof}
\end{eqnarray}
where the expansion is only at ${\cal O}(\Gamma t)^2$ accuracy. It
is similar for $ |A(\overline D^0(t)\to \overline f)|^2$:
\begin{eqnarray}
 |A(\overline D^0(t)\to \overline f)|^2 &=& e^{-\Gamma t} \left[ \overline X_f +\overline Y_f\Gamma t+\overline Z_f (\Gamma
 t)^2+...\right].\label{eq:time-dependentbarDtobarf}
\end{eqnarray}
In the following, we will take $f=f_0K^{*0}$ and $\bar f= f_0\bar
K^{*0}$. If the $K^{*0}(\bar K^{*0})$ meson is reconstructed in
$K^\pm\pi^\mp$ mode, the coefficients in the time-dependent decay
amplitudes squared are given by:
\begin{eqnarray}
 && X_{f_0K^*}=\overline X_{f_0K^*} =|A_{f_0K^*}|^2 r_{ f_0 K^*}^2, \\
 && Y_{ f_0 K^*}=|\frac{q}{p}||A_{ f_0 K^*}|^2r_{ f_0 K^*} (y'_{ f_0 K^*} \cos\phi -x'_{ f_0 K^*} \sin\phi), \\
 &&\overline Y_{ f_0 K^*}=|\frac{p}{q}||A_{ f_0 K^*}|^2r_{ f_0 K^*} (y'_{ f_0 K^*} \cos\phi +x'_{ f_0 K^*} \sin\phi), \\
 && Z_{ f_0 K^*}= |\frac{q}{p}|^2 |A_{ f_0 K^*}|^2 \frac{x^2+y^2}{2}, \label{eq:z} \\
 &&\overline Z_{ f_0 K^*}= |\frac{p}{q}|^2 |A_{ f_0 K^*}|^2 \frac{x^2
 +y^2}{2}.\label{eq:zbar}
\end{eqnarray}
$x'$ and $y'$ are linear combinations of $x$ and $y$:
\begin{eqnarray}
 x_{ f_0 K^*}' &=&(x\cos \delta_{ f_0 K^*}+ y\sin \delta_{ f_0 K^*}), \nonumber\\
 y_{ f_0 K^*}' &=&(y\cos \delta_{ f_0 K^*}- x\sin \delta_{ f_0 K^*}).
\end{eqnarray}
The amplitude squared $|A_{ f_0 K^*}|^2$ can be easily measured
using the time-integrated rate for Cabibbo favored mode $D^0\to
f_0\overline K^{*0}$. In $D^0\to f_0\overline K^{*0}$, the $Y$ and
$Z$ terms come from the mixing followed by the DCS decay $\overline
D^0\to f_0\overline K^{*0}$ which are at least suppressed by
$r_{f_0K^*}\sim 0.04$. Neglecting the power suppressed terms, only
the first term $X=|A_{ f_0 K^*} |^2$ survives and the
time-integrated rate is given by:
\begin{eqnarray}
 \int^\infty_0 |A(D^0(t)\to f_0\overline K^{*0})|^2 dt\approx\frac{|A_{ f_0 K^*} |^2} {\Gamma}.
 \label{eq:D0tof0(980)overlineK*0}
\end{eqnarray}
With $|A_{ f_0 K^*}|^2$ determined,  the ratio $r_{ f_0 K^*}$ is
easily determined by combing the measurements on $X_{ f_0 K^*}$ and
$|A_{ f_0 K^*}|^2$. Furthermore, $|q/p|$ and $x^2+y^2$ are
determined using $Z_{ f_0 K^*}$ and $\overline Z_{ f_0 K^*}$ in
Eqs.~\eqref{eq:z} and \eqref{eq:zbar}   where terms proportional to
$r_{ f_0 K^*}^2$ are neglected.

Using the decay widths given in Eq.~\eqref{eq:time-dependentDtof},
Eq.~\eqref{eq:time-dependentbarDtobarf} and
Eq.~\eqref{eq:D0tof0(980)overlineK*0}, the four parameters are
determined while the other ones ($\phi$, $\delta_{ f_0 K^*}$, $x/y$)
still remain unknown. 
The decay $D^0\to f_0K^{*0}$   where the $K^{*0}$ is identified by
the $K_S\pi^0$ final state could provide more measurements which are
helpful to extract the other parameters. The amplitudes receive two
parts of contributions:
\begin{eqnarray}
 A(D^0\to f_0K^{*}_{ K_S\pi^0})&=&A(\bar D^0\to f_0K^{*}_{ K_S\pi^0})=
 A_{f_0 K^*}[1-r_{f_0K^* } e^{-i\delta_{f_0K^* }}]\equiv A_{f_0K_S\pi},
\end{eqnarray}
where the appropriate reconstruction factors with a $K^{*0}$
identified by the $K_S\pi^0$ are assumed in the experimental
measurements. The coefficients in time-dependent decay widths are
given by:
\begin{eqnarray}
 X_{f_0K_S\pi }&=&\overline X_{f_0 K_S\pi }= |A_{ f_0K_S\pi}|^2, \\
 Y_{f_0K_S\pi }&=&-|\frac{q}{p}| |A_{f_0K_S\pi }|^2(-x\sin\phi+y\cos\phi), \\
 \overline Y_{f_0K_S\pi }&=&-|\frac{p}{q}|
 |A_{f_0K_S\pi }|^2(x\sin\phi+y\cos\phi),
\end{eqnarray}
where we have added the subscript $K_S\pi$ for the coefficients.
With these measurements, the mixing parameters, $\tan^2\phi$ and
$(x/y)^2$, are solved by:
\begin{eqnarray}
 \tan^2 \phi &=& \frac{2f^2-{\cal F}_{ f_0K_S\pi} -\sqrt { {\cal F}^2_{ f_0K_S\pi}-4f^2Y^{(+)2}_{ f_0K_S\pi}}}
               {{\cal F}_{ f_0K_S\pi} +\sqrt { {\cal F}^2_{ f_0K_S\pi}-4f^2Y^{(+)  2}_{ f_0K_S\pi}}},  \label{eq:tanphisolution}\\
  \frac{x^2}{y^2}&=&\frac{{\cal F}_{ f_0K_S\pi} -2Y^{(+)2}_{ f_0K_S\pi}+\sqrt { {\cal F}^2_{ f_0K_S\pi}-4f^2Y^{(+)2}_{ f_0K_S\pi}}}
               {2Y^{(+)2}_{ f_0K_S\pi}},\label{eq:x^2/y^2solution}
\end{eqnarray}
where $\cos 2\phi$ is chosen positive (very small $\phi$ in SM). For
convenience, the two coefficients have been reexpressed by:
\begin{eqnarray}
 Y^{(+)}_{ f_0K_S\pi}&=&\frac{\overline Y_{ f_0K_S\pi}|q|^2+Y_{ f_0K_S\pi}|p|^2}{2X_{ f_0K_S\pi}|q||p|}=-y\cos\phi, \\
 Y^{(-)}_{ f_0K_S\pi}&=&\frac{\overline Y_{ f_0K_S\pi}|q|^2-Y_{ f_0K_S\pi}|p|^2}{2X_{ f_0K_S\pi}|q||p|}=-x\sin\phi,
\end{eqnarray}
and ${\cal F}_{ f_0K_S\pi}= f^2+ Y^{(+)2}_{ f_0K_S\pi}- Y^{(-)2}_{
f_0K_S\pi}$. In the limit of $\phi\to0$, the measured value of
$Y^{(-)}_{ f_0K_S\pi}$ is very small and thus we can expand the
above solutions in Eq.~\eqref{eq:tanphisolution} and
Eq.~\eqref{eq:x^2/y^2solution} into:
\begin{eqnarray}
 \tan^2 \phi &=& \frac{Y^{(-)  2}_{ f_0K_S\pi}}
               {f^2-Y^{(+)  2}_{ f_0K_S\pi}}+..., \label{eq:tan^2phiexpansion} \\
  \frac{x^2}{y^2}&=&\frac{f^2-Y^{(+)  2}_{ f_0K_S\pi}}{Y^{(+)2}_{ f_0K_S\pi}}
              - \frac{Y^{(-)2}_{ f_0K_S\pi}f^2}{Y^{(+)2}_{ f_0K_S\pi}(f^2-Y^{(+)
              2}_{ f_0K_S\pi})}+...,\label{eq:x^2/y^2expansion}
\end{eqnarray}
where higher powers of $Y^{(-)}_{ f_0K_S\pi}$ were neglected. As we
can see from Eq.~\eqref{eq:x^2/y^2expansion}, there are two
advantages in our method: the ratio of $x^2$ and $y^2$ is finite
even for tiny $\phi$; our method does not need the strong phase
difference $\delta$ which is unknown at all.

The authors in Ref.~\cite{Sinha:2007ck} propose to extract the
mixing parameters using the time-dependent study of $D^0\to
K^{*0}\pi^0$. The four parameters, $|A_{ f_0(\pi^0) K^*}|^2$, $|A_{
f_0(\pi^0) K^*}|^2$, $|p/q|$ and $f^2=x^2+y^2$, are determined in
the similar way but the other parameters are extracted in different
ways. In $D^0\to K_S\pi^0\pi^0$ decays, the final state contains two
neutral pion. It requires four photons to reconstruct them which is
a rather difficult job. Thus they  suggest to use the normalization
of $D^0\to K_S\pi^0\pi^0$ to give a constraint on the strong phase
difference. In the present method, the normalization constraint is
not used and we utilize measurements on the time-dependent decay
width of $D^0\to f_0K^{*}_{K_S\pi^0}$ instead. Although the $D^0\to
f_0K^{*}$ has a small branching ratio, the advantage is that the
$f_0$ meson is easier to re-construct than $\pi^0$ on the
experimental side.

In summary, we have studied the $D^0-\overline D^0$ mixing in $D\to
f_0K^{*}$ decay,   where the $K^{*0}$ meson is reconstructed in both
$K^+\pi^-$ and $K_S\pi^0$ final state. The method can be directly
generalized to the three-body decay $D\to \pi^+\pi^- K^*$ where the
$\pi^+\pi^-$ is an $S$-wave state. The $f_0$ meson can also be
replaced by a longitudinally polarized $\rho^0$ or $\omega$ meson.
All parameters in $D^0-\overline D^0$ mixing can be extracted by
studying the time-dependent decay widths of these channel. The
extraction is valid in all regions for mixing parameters and it does
not depend on the strong phase difference defined in
Eq.~\eqref{eq:ratio}. The future experimental studies on $D\to
f_0K^{*}$ decay, together with the $D\to \rho^0(\omega)K^*$ decays
in which  vector mesons are longitudinally polarized and the
three-body decay $D\to \pi^+\pi^- K^{*}$, can provide another
alternative method to extract the mixing parameters.

{$Acknowledgements$}: We would like to thank Hai-Bo Li, Rahul Sinha,
Mao-Zhi Yang and Shun Zhou for valuable discussions. This work is
partly supported by National Nature Science Foundation of China
under the Grant Numbers 10735080 and 10625525.

\end{document}